\documentclass[10pt]{article}
\usepackage{marvosym}
\usepackage{graphicx}
\usepackage{authblk}
\usepackage{titlesec}

\newcommand{\Keywords}[1]{\noindent \textbf{keywords:} #1}

\title{\uppercase{Studying the topological stability of the $\lambda\Phi^4 $ kink.}}

\author[1]{Hern\'an Piragua}
\author[2]{Patricio S.  Letelier}

\affil[1]{Instituto de F\'isica, Universidade Estadual de Campinas, CEP 13083-970, Campinas, SP, Brazil, hpiragua@ifi.unicamp.br}
\affil[2]{Departamento de Matem\'atica Aplicada, Universidade Estadual de Campinas, CEP 13083-859, Campinas, SP, Brazil, letelier@ime.unicamp.br}

\begin{document}

\maketitle

\begin{abstract}
The lambda-phi4 kink is linearly and topologically stable. We  study how  extra energy  perturbations are dissipated beyond the linear regime. We found that depending on  the width, amplitude and energy of a Gaussian perturbation  different scenarios are possible: radiation, oscillons creation, kink anti-kink pairs production and shock waves. 
\end{abstract}
\Keywords{Applications of Nonlinear Sciences, Nonlinear Dynamics, kink, stability, lambda phi 4.} \\

\section{Basic information}
Scalar fields with a $\lambda\Phi^4 $ potentials  has been studied along several decades, it has been used to model the behavior  of physical systems in several branches of physics like condensed matter, plasma physics and field theory among others. This toy model is interesting because although it is one of the  of simplest models, it is non-integrable \cite{mcleod}. This model has a kink and a anti-kink solution, see \cite{rajaraman}.  The kink is linear and topologically stable, meaning that it is a  kind of `indestructible' solution of the field equations. In the mid 70's it was found numerically, that after the collision of a kink with an antikink, something like a breather solution appeared, see \cite{kosevich}. Segur and Kruskal showed that a breather is unable to exist \cite{segur}, but it is  an unstable oscillatory solution of the of the $\lambda\Phi^4$ theory,
 that  has an unusual long time existence. After that article, new work have been done studying this oscillon, its existence, stability (see for instance  \cite{gleiser}) and its interaction with kinks.

The aim  of the present paper is  to  revisit the behavior of the    of the 
$\lambda\Phi^4$ kink under perturbations. We know that it is stable, but we want to know how extra energy of a perturbation is dissipated. It is well known that when the perturbation is small enough, the extra energy is dissipated through radiation. Our interest is to study beyond this regime, increasing the amplitude beyond the linear domain and further if possible.

\section{The model}
The Lagrangian density of the model is,
\begin{equation}
 \mathcal{L}=\partial_{\mu}\phi\partial^{\mu}\phi-U(\phi),
\end{equation}
with the potential $U(\phi)=1/4(\phi^2-1)^2$. The equation of motion in one dimension is,
\begin{equation}
 \partial^2_{t}\phi-\partial^2_{x}\phi=U'(\phi)=\phi^3-\phi.
\label{eqmov}
\end{equation}
We use units such  that $c=1$. This equation has the static solutions,
\begin{equation}
 \phi(x)=\pm\tanh\left(\frac{1}{\sqrt{2}}(x-x_k)\right),
\end{equation}
that are  known as  kinks ($+$) and anti-kinks ($-$).

 We shall  perturb the kink with a perturbation  of Gaussian profile. This profile  is chosen because it is smooth and localized. In this way, our perturbation  problem can be defined as solving equation (\ref{eqmov}) with the initial condition,
\begin{equation}
 \phi(x,0)=\tanh\left(\frac{1}{\sqrt{2}}x\right)+A e^{-b(x-x_0)^2},
\end{equation}
where the constants $A,b$ and $x_0$ describe the amplitude, inverse of width and position of the Gaussian perturbation.

\section{The method and preliminary results}

To solve our differential equation we use a finite difference scheme, for simplicity we use the Leap-Frog scheme that is of the  order two in time and space and it is also explicit, then  it is not difficult  to implement. Fixed boundaries far enough from the kink and the perturbation are imposed. We note that three parameters can be tuned in the perturbation $A,b$ and $x_0$.

\begin{figure}[p]
    \centering
 \includegraphics[width=120mm]{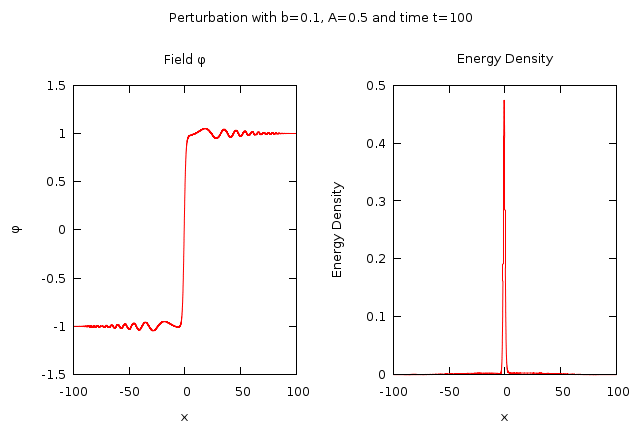}
\caption{With a small initial amplitude, all energy dissipates in radiation.}
\label{fig1}
\end{figure}

\begin{figure}[p]
    \centering
 \includegraphics[width=120mm]{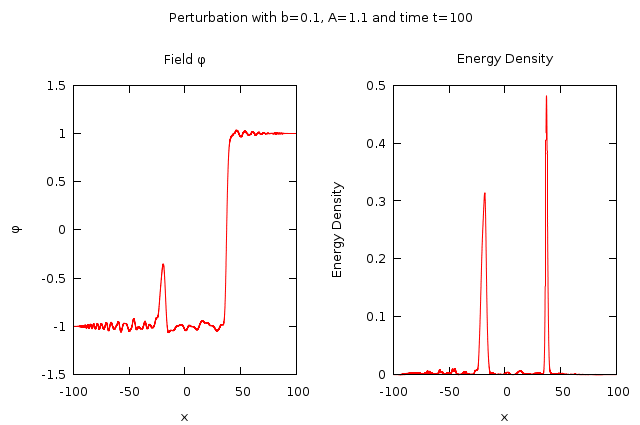}
\caption{If the amplitude is increased, an oscillon is formed.}
\label{fig2}
\end{figure}

\begin{figure}[p]
\centering
\includegraphics[width=120mm]{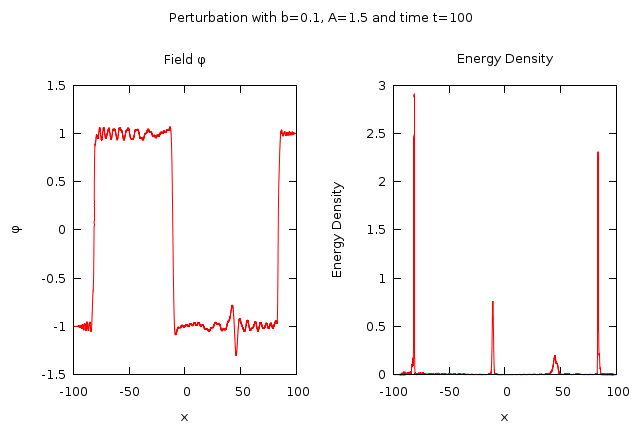}
\caption{When energies go over four times the kink energy, kink-anti kinks can be created.}
\label{fig3}
\end{figure}

Different results are obtained depending on the amplitude ($A$) and the width ($1/b$). When $b$ is of the order of $0.1$ and the amplitude is less than one, basically all the perturbations are transformed  in radiation. This behavior is shown in figure \ref{fig1}. When we further increase the amplitude, an oscillon is created, see figure \ref{fig2}. Beyond the threshold of four times the energy of the kink, it can appear a kink-anti kink pair and if the energy is high enough an oscillon, see figure \ref{fig3}. 

\begin{figure}[p]
    \centering
 \includegraphics[width=120mm]{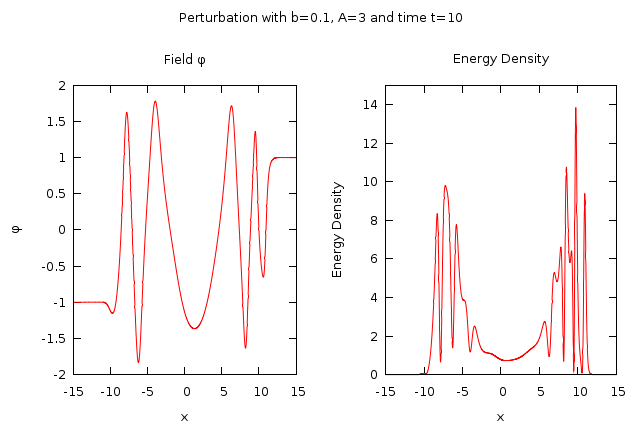}
\caption{Almost $70$ times the kink enery, it shows oscillation around $\phi=0$, the double well is unnoticed.}
\label{fig4}
\end{figure}

\begin{figure}[p]
    \centering
 \includegraphics[width=120mm]{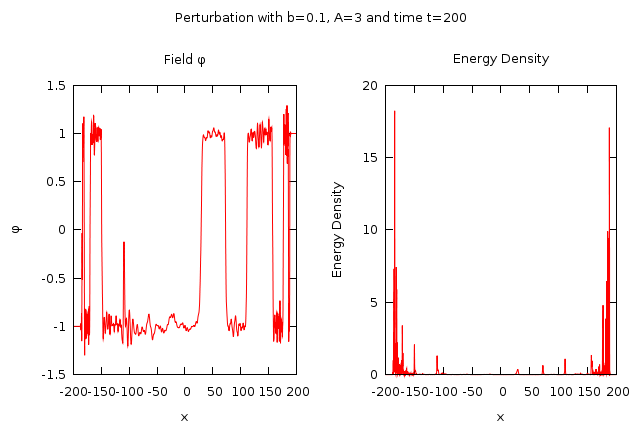}
\caption{Almost $70$ times the kink enery, Some time later, the creation of structures begin.}
\label{fig5}
\end{figure}

If the amplitude is further increased, the double well is ignored and only the $ \lambda\Phi^4 $ potential is seen. In this case it is observed oscillations around $\phi=0$. See for example figure \ref{fig4}, where $A=3$ and the total enery of the system is almost $70$ times the energy of the kink. Then, the perturbation is spread, its amplitude decreases while going away from the kink. When the field is of the same order of unity, creation of structures occurs, see figure \ref{fig5}.

\begin{figure}[p]
    \centering
 \includegraphics[width=120mm]{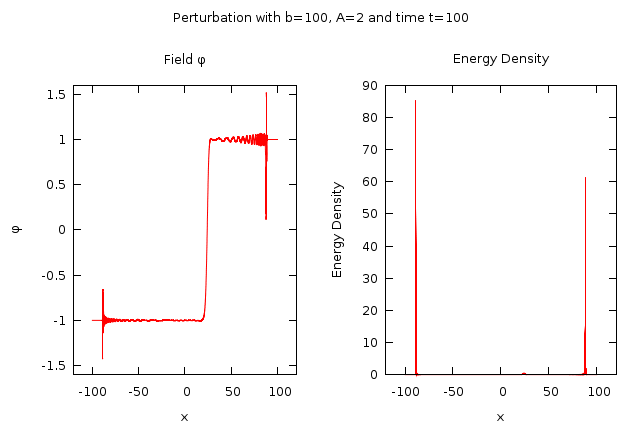}
\caption{When the perturbation is very sharp, $b\sim100$.}
\label{fig6}
\end{figure}

If $b$ is smaller, $b\sim 0.01$, i.e. the perturbation is broader similar results are obtained. But the formation of oscillons at lower energies is more likely. For $b$ high, $b\sim100$, the perturbation is very sharp and it seems that the energy goes away from the kink like shock waves, see figure \ref{fig6}.

\begin{figure}[p]
    \centering
 \includegraphics[width=150mm]{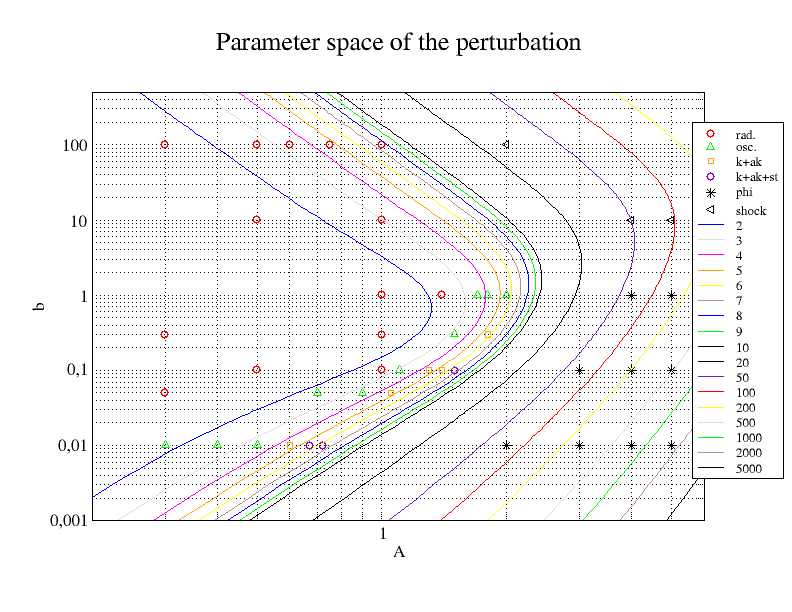}
\caption{When the perturbation is very sharp, $b\sim100$.}
\label{fig7}
\end{figure}

It is very interesting the different behaviors of the field under different perturbations. In figure \ref{fig7} we show the parameter space of the perturbation. The curves are equal energy lines, the legend shows the value in terms of the kink energy, i.e. $2$ means total energy twice the kink. Also this figure shows the outcome of the perturbation at the time $t=100$ units. ``rad.'', means radiation, ``osc.'' an oscillon was created, ``k+ak'' a kink anti-kink pair, ```k+ak+st'' a kink anti-kink plus an oscillon or two, ``phi'' means that the double well is unnoticed and finally ``shock'' stands for a lot energy is in the border, like figure \ref{fig6}  For a given amplitud we can obtain the same energy with two widths. If is sharp enough, the spatial derivative term in the kinetic energy is huge, and on the opposite case we have a lot of energy because of the bigger area of the perturbation.
\section{Discussion}
We did not find in the literature where  large perturbations  were explored.
Due to nonlinear character of the field equation, some precautions have to be taken, some regions will oscillate rather quickly in time and space and it is important to have a small grid spacing. But, there are  zones where this small grid  spacing  is not needed. There are schema  like the AMR (Adaptive Mesh Refinement), where the refinement is applied intelligently to adjust itself and reduce calculations.

\section{Acknowledgments}
We  thank FAPESP for financial support and also P.S.L thanks CNPq for partial financial support.


\end{document}